\documentclass{aastex631}
\usepackage{etoolbox}

\begin{document}

\title{GRB 221009A: Spectral signatures based on ALPs candidates}

\correspondingauthor{D. Avila Rojas}
\email{daniel\_avila5@ciencias.unam.mx}

\author[0000-0002-4020-4142]{D. Avila Rojas}
\affiliation{Instituto de Física, Universidad Nacional Autónoma de México, Ciudad de Mexico, 04510, Mexico}

\author[0000-0002-2565-8365]{S. Hernández-Cadena}
\affiliation{Instituto de Física, Universidad Nacional Autónoma de México, Ciudad de Mexico, 04510, Mexico}

\author[0000-0002-5209-5641]{M. M. González}
\affiliation{Instituto de Astronomía, Universidad Nacional Autónoma de México, Ciudad de Mexico, 04510, Mexico}

\author[0000-0002-8940-5316]{A. Pratts}
\affiliation{Instituto de Física, Universidad Nacional Autónoma de México, Ciudad de Mexico, 04510, Mexico}

\author{R. Alfaro}
\affiliation{Instituto de Física, Universidad Nacional Autónoma de México, Ciudad de Mexico, 04510, Mexico}

\author[0000-0002-8774-8147]{J. Serna-Franco}
\affiliation{Instituto de Astronomía, Universidad Nacional Autónoma de México, Ciudad de Mexico, 04510, Mexico}

\begin{abstract}

GRB 221009A has posed a significant challenge to our current understanding of the mechanisms that produce TeV photons in gamma-ray bursts (GRB). On one hand, the Klein-Nishina (KN) effect of the inverse Compton scattering leads to less efficient energy losses of high-energy electrons. In the other hand, at a redshift of 0.151, the TeV spectrum of GRB 221009A undergoes significant absorption by the Extragalactic Background Light (EBL). Therefore, the observation of a 13-TeV photon in this event implies the presence of enormous photon fluxes at the source, which the Synchrotron Self-Compton mechanism in external shocks cannot easily generate. As an alternative, some authors have suggested the possibility of converting the TeV-photons into Axion-like particles (ALPs) at the host galaxy, in order to avoid the effects of EBL absorption, and then reconverting them into photons within the Milky Way. While this solution relaxes the requirement of very high photon fluxes, the KN effect still poses a challenge. Previously, we have shown that the injections of ALPs could explain the observation of 13-TeV photons. Here, we include the energy dependence of the survival probability and the amount of energy carried to determine the ALP candidates which could potentially explain the TeV photons observed by LHAASO and their hard spectrum. We found that the allowed candidates are generally cluster around masses of $10^{-7}$ eV. We also considered different EBL models, for the one predicting larger attenuation tends to reject ALP candidates with the lowest coupling factor. For some hypothesis of EBL model, these candidates are found below a region of the parameter space in which, if detected, ALPs could account for all of the cold dark matter in the Universe.

\end{abstract}

\keywords{gamma-ray burst: general --- gamma-ray burst: individual (221009A)  --- gamma rays: general --- emission processes --- dark matter --- axion-like particles --- ALPs   }

\section{GRB 221009A observations} \label{sec:intro}

GRB 221009A was first detected on  2022 October 9th by the Fermi-GBM \citep{2022GCN.32636....1V} and was followed by several other instruments in many different wavelengths \citep{2022GCN.32632....1D,2022GCN.32680....1T,2022GCN.32852....1P,2022GCN.32644....1H,2022GCN.32655....1F, 2022GCN.32767....1L,2022GCN.32654....1D,2022GCN.32652....1B}.  After preliminary analysis of the first observations, $T_{90}$, fluence and isotropic energy for this event were derived, making this GRB the most luminous ever detected \cite{2022GCN.32642....1L,2018ApJ...869L..23T}. Spectroscopic observations from GTC's OSIRIS and the VLT's X-Shooter were used to estimate the corresponding redshift of $z=0.151$ \citep{2022GCN.32686....1C, 2022GCN.32648....1D}. Observations from the Hubble Space Telescope (HST), reveal properties of the host galaxy confirming that the galaxy is prototypical of host galaxies associated with long GRBs. HST observations also show that this GRB is located near the center of its host galaxy \citep{Levan_2023}.

 There is moderate statistical evidence for a supernova associated with GRB221009A \citep{2023ApJ...946L..25S,2023ApJ...949L..39S}. The James Web Space Telescope analyzed the afterglow emission and found little evidence for variability from early to late times, implying modest contributions from supernova emission. Furthermore, the extreme properties of the burst are likely not linked to an extreme and unusual environment \citep{Levan_2023}.

High energy emission was reported by Fermi-LAT and the Large High Altitude Air Shower Observatory (LHAASO) in the energy range between GeV and TeV. The afterglow temporal profile was fitted from Fermi-LAT data as a power law with an index of $1.32\pm 0.05$ \citep{Levan_2023}. A search for very-high-energy (VHE) photons with no significant detection from this GRB was performed by the High Altitude Water Cherenkov (HAWC) observatory and the  High Energy Stereoscopic System (H.E.S.S.) 8 and 53 hours after the trigger, respectively, resulting in flux upper limits \citep{2023ApJ...946L..27A}. In particular, the H.E.S.S. upper limit for energies above 650 GeV rules out the inverse Compton scenario, at least at those times, with the X-ray emission as the synchrotron counterpart. The Fermi Collaboration has published its first results for the light curve using data from Fermi-GBM and Fermi-LAT. The total fluence and isotropic energy between $1-10,000\,\rm{keV}$ derived from the individual time intervals from $t_{0}-2.7\,\rm{s}$ to $t_{0}+1449.5\,\rm{s}$ in the Fermi data are $S=(9.47\pm 0.07)\times 10^{-2}\,\rm{erg}\,\rm{cm^{-2}}$ and $E_{\rm{iso}}= (1.01\pm 0.007)\times 10^{55}\,\rm{erg}$, in accordance with other instruments \citep{2023arXiv230314172L,2023arXiv230300898Y}. Fermi-LAT early emission reported a photon with energy of 99.3 GeV at 240 s after GBM trigger \citep{2022GCN.32658....1P} and posterior analysis found the highest photon detected by LAT with 397.7 GeV at 33554 s after GBM trigger \citep{2022GCN.32748....1X}. 

One of the highlights from GRB 221009A was the very first detection, performed by LHAASO, of VHE photons with energies $>10\,\rm{TeV}$ from a GRB \citep{2022GCN.32677....1H}. The detection of at least one photon above $10\,\rm{TeV}$, considering the redshift measured for this GRB, is not easily explained by the synchrotron-self Compton (SSC) scenario. Depending on the Extragalactic Background Light (EBL) model, an attenuation larger than $10^{-3}$ is expected for photons with energies above $10\,\rm{TeV}$, resulting in fluxes of $\mathcal{O}(10^{-9}\, \text{TeV}^{-1}\, \text{cm}^{-2}\, \text{s}^{-1})$. These fluxes would be too low for even LHAASO-KM2A to detect a single photon with an energy of 13 TeV if only the SSC contribution is assumed. Similar results have been obtained by several authors as \citep{2023MNRAS.522L..56S,2023ApJ...947L..14Z,2023ApJ...947...53R} including the LHAASO collaborations \citep{2023Sci...380.1390L, sciadv.adj2778}.

There was a first publication using only data taken by the WCDA instrument \citep{2023Sci...380.1390L}. The detected number of photons reached over 64,000 within the first $\sim3000$ seconds. The intrinsic and observed spectra were presented in the limited energy range of 0.2 - 7 TeV (its sensitivity extends up to energies of 100 TeV), showing no indication of a spectral break or cutoff in the intrinsic spectrum up to 5 TeV.  Later, a second publication with data from KM2 \citep{sciadv.adj2778} reported more than 140 photons detected with a significance of $20.6\sigma$ above an energy of 3 TeV and between $230$ to $900$ seconds after GBM trigger. The maximum energy estimated depends on the spectral model assumed, being $17.8^{+7.4}_{-5.1}\,\text{TeV}$ with a log-parabola and $12.2^{+3.5}_{-2.4}\,\text{TeV}$ with a power-law with an exponential cutoff. There is also a claim of a new spectral component above energies of 7 TeV, in addition to the SSC contribution, not explained by conventional physics. Because of the difficulties in explaining this very-high-energy spectral component, the authors explored the axion particle scenario, confirming most of the already excluded region in the parameter space of the axion coupling factor and axion mass by the Supernova SN1987A \citep{2015JCAP...02..006P}, Fermi \citep{2016PhRvL.116p1101A,2021PhLB..82136611C} and H. E. S. S. \citep{2019A&A...627A.159H}. They added to the excluded region a small zone around masses of $10^{-7}$ eV and coupling factors of $3\times 10^{-11} \rm GeV^{-1} $.

Another interesting observation of this GRB is the reported detection by Carpet-2 of a 251-TeV photon 4563 s after GBM's trigger \citep{ATel15669}. While certainly intriguing, it is more likely that this photon is associated with a Galactic source as pointed out by the HAWC Collaboration \citep{2022ATel15675....1F}. For this reason, in this paper, we only focus on the VHE photons reported by LHAASO. 

The origin of the VHE emission from GRBs has been explored by assuming different scenarios. The most common one is the SSC model, in which VHE gamma rays are generated in the external shocks of the GRB jet. It is possible to reach energies up to $1\,\rm{TeV}$ from the forward shock emission by assuming that the electron spectrum has no energy cutoff \citep{2023ApJ...946L..23L}. 
Then, even if the jet structure is assumed to be more complex to reach such energies, such a high value of the differential flux would have profound implications for the energy budget of the event \citep{2023ApJ...946L..24W,2023MNRAS.tmpL..35S}. Therefore, so far, the TeV emission from GRB 221009A is not well described by standard SSC models \citep{2023arXiv230206225K} although other multifrequency detections of the afterglow emission are completely in agreement with SSC.

There have been several searches for neutrinos correlated to GRBs by instruments such as IceCube \citep{2017ApJ...843..112A,2021ApJ...910....4A,2022ApJ...939..116A,2023arXiv231211515I}, Amanda \citep{2007ApJ...664..397A,2008ApJ...674..357A} and Antares \cite{2017MNRAS.469..906A,2021MNRAS.500.5614A,2021JCAP...03..092A} that have resulted in no detections and in strong constraints on the single-zone fireball models of neutrinos and the production of Ultra-high Energy Cosmic Ray (UHECR) during the precursor, prompt and afterglow phases \citep{2023A&A...672A.102L}. Despite being considered likely sources of UHECRs due to their large power output, the results of searches for neutrinos correlated with GRBs have been discouraging. As a result, attention has shifted towards other sources, such as active galactic nuclei.

IceCube conducted searches for neutrinos at various time ranges, hours, and days from the initial Fermi-GBM trigger, between $800\,\rm{GeV}-1\,\rm{PeV}$. However, no events were found to be coincident with the position of GRB 221009A \citep{2022GCN.32665....1I,2023ApJ...946L..26A}. Since the non-detection of neutrino emission, the models that forecast the production of neutrinos from GRBs \citep{2022arXiv220206480K} were evaluated by testing the lack of detection. The derived upper limit allowed for the refinement of GRB model parameters \citep{2022ApJ...941L..10M}. The absence of detection led to a diligent effort to explain this situation through the exploration of various prompt models, as described by \citep{2023ApJ...944..115A,2023ApJ...944L..34R}, and precursor models \citep{2023MNRAS.521.2391B}. Furthermore, novel models that circumvent the EBL attenuation due to heavy neutrinos emerged as a prospective solution \citep{2022arXiv221014178C,2022arXiv221100634S}.

Preceding the publication on LHAASO-KM2A \citep{sciadv.adj2778}, some models were proposed assuming the acceleration of UHECR in this GRB. For instance, UHECR from internal shocks could interact with the EBL and generate photons with energies up to $10\,\rm{TeV}$ \citep{2023ApJ...944L..34R}. It has also suggested that UHECR from the reverse shock could also produce 18-TeV photons (as initially reported) via synchrotron radiation, assuming an optimistic value of the coefficient $\eta = 1$ in the proton acceleration timescale \citep{2022arXiv221105754Z}. However, for relativistic and ultra-relativistic shocks as expected in GRBs, the coefficient is greater than unity \citep{2015SSRv..191..519S}. If a more conservative value of the coefficient ($\eta = 10$) is used, it becomes challenging to detect photons with energies $> 10\,\rm{TeV}$. Finally, if the GRB is situated at the outer regions of its host galaxy and the Extra-galactic Magnetic Field (EGMF) is as low as $\mathcal{O}(10^{-14}\,\rm{G})$ (to decrease the time delay due to UHCR propagation below 2000 s), photons from the particle cascades of UHECR could explain the LHAASO detection \citep{2023A&A...670L..12D}.

Some other studies have proposed that an 18-TeV photon detection can be explained by Lorentz Invariance violation effects on  $\gamma - \gamma$  absorption \cite{2023APh...14802831L,2023ApJ...942L..21F,2023PhRvD.107h3001Z} or non-standard physics such as the existence the Axion Like Particles (ALPs). It was proposed in reference \cite{2022arXiv221005659G} the conversion of TeV-photons into ALPs at the host galaxy to avoid the effects of EBL absorption, followed by reconversion into photons within the Milky Way. In reference \cite{2022arXiv221007172B}, a parameter space of ALP candidates was explored to identify those with the highest probability of arriving on Earth as photons. A bound on the Lorentz boost was also obtained. Similarly, reference \cite{2022JETPL.116..767T} explored the parameter space of ALPs, but only by considering conversions to photons due to the magnetic field of the Milky Way. The authors assumed that the flow of ALPs reaching the Milky Way constitutes one-third of the total flux of photons emitted by the GRB. While these solutions relax the requirement of very high photon fluxes, the KN effect still poses a challenge. In a previous publication \citep{2023ApJ...944..178G}, we considered an initial injection of ALPs and calculated the probability of their arrival on Earth as TeV photons. Our results show a region of allowed parameters of ALPs that could potentially explain the detection of the 18-TeV photon by the LHAASO observatory. Our results stand even when considering a lower maximum energy of the emission of 13-TeV as later published by LHAASO-KM2A. 
In this paper, we extend our analysis to include the energy dependence of the survival probability of ALPs and the amount of energy carried, in order to determine the spectral conditions that allow for the detection of 13-TeV photons while avoiding the detection of photons with energies greater than 25 TeV (section \ref{sec:spectra}).


\section{Spectral signatures from ALPs} \label{sec:spectra}

ALPs are hypothetical particles that arise in theories beyond the standard model and are related to the Peccei-Quinn mechanism \citep{1977PhRvL..38.1440P}. This mechanism was proposed to solve the strong CP problem in Quantum Chromodynamics (QCD) and involves the breaking of a symmetry (PQ symmetry) that generates a particle called the Axion \citep{PhysRevLett.40.223,PhysRevLett.40.279,PhysRevLett.43.103}. Thus, ALPs are an extension of the concept of Axion that does not requires to solve the strong CP problem and are considered as candidates for cold dark matter (see \cite{IRASTORZA201889} for a review). The Lagrangian of ALPs is given by

\begin{equation}
    \mathcal{L}_{a} =\frac{1}{2}(\partial_{\mu}a \partial^{\mu}a - m^{2}_{a}a^{2}) + \frac{1}{4f_{a}}aF_{\mu \nu}\tilde{F}^{\mu \nu}. 
\end{equation}

\noindent where $a$ is the ALP field , $m_{a}$ is the  ALP mass , $F_{\mu \nu}$ is the  Faraday Tensor and $\tilde{F}^{\mu \nu}$ is its dual. ALPs are of special interest because they are coupled to electromagnetism, which allows for oscillations between photons and ALPs under the influence of source, galactic, and intergalactic magnetic fields. The coupling is expressed as:

\begin{equation}
 \mathcal{L}_{a\gamma} = \frac{1}{4f_{a}}aF_{\mu \nu}\tilde{F}^{\mu\nu}= ag_{a \gamma} \vec{E}\cdot\vec{B},
\end{equation}
 
\noindent where  $g_{a \gamma}$ is the coupling constant and $\vec{E}$, $\vec{B}$ are the electric and magnetic fields respectively \citep{PhysRevD.37.1237}. An ALP with a kinetic energy in the TeV range can undergo oscillations and transform into a TeV-photon at any point during its journey towards Earth \citep{Mirizzi_2017}, to survive the attenuation due to the EBL, and reach the TeV detectors. The propagation of the ALP-photon system along the line of sight is described by the propagation equation given by \citep{PhysRevD.84.105030}. 

In this study, we do not make any assumption on the existence of an initial population of photons due to their inefficient production by the preferred mechanism, the SSC. Moreover, if TeV photons are produced, their flux must be very high to compensate for the attenuation caused by the EBL. As mentioned in section \ref{sec:intro}, some authors proposed the possibility of photons transforming into ALPs to avoid EBL attenuation, but this scenario would also require a high photon flux, as not all photons would necessarily oscillate into ALPs in the host galaxy or the GRB jet, and not all that do oscillate would convert back to photons in the Milky Way. Moreover, the claim of an extra spectral component by LHAASO justifies our hypothesis of considering an initial beam consisting solely of ALPs, with the density matrix, $\rho = \Phi \Phi ^{\dag}$, at $t=0$ represented as $\rho(0)= diag (0,0,1)$. The probability of survival, $P_{a\gamma}$, resulting from ALPs being detected on Earth as photon is calculated by incorporating all physical parameters, such as those for the medium, magnetic field, electron density, and propagation distance, into the mixing matrix $\mathcal{M}$. The evolution equation for the density matrix is given by,

\begin{equation}
    i \frac{\partial \rho}{\partial l}= [\rho , \mathcal{M}],
\end{equation}

\noindent where $l$ is the propagation distance for a given medium. The survival probability is then,
\begin{equation}
    P_{a\gamma}=\rho_{1,1}(l) + \rho_{2,2} (l) ,
    \label{survalp}
\end{equation} where $\rho_{1,1}$, $\rho_{2,2}$ represent the first and second diagonal elements of the density matrix  \citep{Bi_2021} .

We use the open code \texttt{gammaALP}\footnote{\url{https://gammaalps.readthedocs.io/en/latest}}, that allow us to calculate the survival probability taking into account different astrophysical environments as Galaxy Host, EBL attenuation, and the Milky Way \citep{Meyer_2021}. Table \ref{tab:param} shows the parameter values used to calculate the survival probability using \texttt{gammaALP}. We consider the Gilmore \citep{Gilmore_2012} EBL model, which predicts a moderate attenuation. Later, we discuss the impact of two extreme EBL models on our results.

\begin{table}[t]
\centering
\begin{tabular}{lcc}
\hline
Parameter & Value & Reference\\\hline
$E_{\rm iso}$ & $1.01\times10^{55}$ erg &  \citep{2023arXiv230314172L}\\
$d_{z}$ & $2.23\times10^{27}$ cm &\\
EBL model & Gilmore& \citep{Gilmore_2012}\\
$B_{\rm MW}$  &  $\sim 3 \mu   $G & \citep{Jansson_2012} \\
$B_{\rm HG}$ & $\sim 3 \mu  $G & set as MW \\
$d_{\rm HG}$ & 30 $\rm kpc$ & set as MW\\
$d_{\rm MW}$ & $\rm < 30 kpc$ & $\leq d_{\rm MW}$\\
$\alpha_{\rm a}$ & $2.6,~2.8,~3.0$ & set\\
$E_{\rm DM}$ & $0.1,~1,~3\%$ & set\\ \hline
\end{tabular}
\caption{\textbf{Parameters considered in our calculations for the photon survival probability from ALPs-photon conversions. Here, $E_{iso}$ corresponds to the isotropic total energy of the burst, $d_{z}$ stands for the luminosity distance of the GRB, $B_{\rm MW},~B_{\rm HG}$ are the Milky Way and Host Galaxy magnetic fields respectively, $d_{\rm MW},~d_{\rm HG}$ are the disk diameter of the Milky Way and Host Galaxy respectively, $\alpha_a$ is the ALP spectral index, and $E_{DM}$ is the percentage of energy carried by the ALP population.}}
\label{tab:param}
\end{table}

Previously in \cite{2023ApJ...944..178G}, we assumed an injection of ALPs with a spectrum that depended solely on the ALP energy, $E_{\rm a}$, and was described by a power law function. The values considered for the ALP spectral index, $\alpha_{a}$, were $1.8$ and $2.5$. Here, we introduce a temporal profile, $\tau(t)$, to the injection while keeping the energy dependency. Then, the ALP spectrum is given by,

\begin{equation}
\Phi_{\mathrm{a}}(E_{\mathrm{a}},t) = N_{o}\tau(t)\left(\frac{E_{\mathrm{a}}}{E_{a,0}}\right)^{-\alpha_{\mathrm{a}}}\text{,}
\end{equation}

where $E_{a,0}=1\,\rm{TeV}$. We consider different values of the spectral index $\alpha_{a}$ between $2.0$ and $3.0$, along with the temporal profile given by LHAASO-KM2A detection \citep{sciadv.adj2778}. Thus, we obtain $N_{o}$ by considering that the energy carried by the ALPs is a fraction of the total GRB's isotropic energy. Then,

\begin{equation}
\%E_{\rm iso} = \frac{4\pi d_{z}}{1+z}N_{o}\int_{E_{\rm a,min}}^{E_{\rm a,max}}\int_{t_{\rm min}} ^{t_{\rm max}}\tau(t)\left(\frac{E_{\rm a}}{1\,\rm{TeV}}\right)^{-\alpha_{\rm a}+1}dtdE_{\rm a}\text{,}
\label{eqn:Eiso}
\end{equation}

\noindent where $d_{z}$ is the luminosity distance to the source, $z$ is the redshift, $t_{\rm{min}}=300\,\rm{s}$ and $t_{\rm{max}}=900\,\rm{s}$ define the time interval when the ALP injection occurs and the 13-TeV photons are detected, and $E_{\rm a,min}$ and $E_{\rm a,max}$ are the maximum and minimum energies of the injected ALPs. We consider that the total energy carried by the ALPs could represent from 0.1 up to 3\% of $E_{\rm{iso}}$. Finally, the number of observed photons by LHAASO, $N_{\gamma}$, is obtained by considering the effective area $A(E_{\gamma})$ of LHAASO-KM2A as function of the photon energy ($E_{\gamma}$), the normalization ($N_{o}$) of the ALPs spectrum, and the survival probability, $P_{a\gamma}(E_{\gamma})$, and is given as,

\begin{align}
N_{\gamma} &= \int_{E_{\gamma,{\rm min}}}^{E_{\gamma,{\rm max}}}\int_{t_{\rm min}}^{t_{\rm max}}\Phi_{\gamma}(E_{\gamma}) A(E_{\gamma})dtdE_{\gamma} \\
&= N_{o} \int_{E_{\gamma,{\rm min}}}^{E_{\gamma,{\rm max}}}\int_{t_{\rm min}}^{t_{\rm max}}\Phi_{\rm a}(E_{\rm a},t) P_{{\rm a}\gamma}(E_{\gamma})A(E_{\gamma})dtdE_{\gamma} \\
&= N_{o} \int_{E_{\gamma,{\rm min}}}^{E_{\gamma,{\rm max}}}\int_{t_{\rm min}}^{t_{\rm max}}\tau(t)\left(\dfrac{E_{\rm a}}{1\,\rm{TeV}}\right)^{-\alpha_{\rm a}}P_{{\rm a}\gamma}(E_{\gamma})A(E_{\gamma})dtdE_{\gamma}\text{.} \label{eqn:obsph}
\end{align}

Equation \ref{eqn:obsph} is solved considering equation \ref{eqn:Eiso}, that the injection of ALPs last as long as the photon observation and, that $E_{\rm a}\sim E_{\gamma}$ since the mass of the ALP is much smaller than its kinetic energy. Note that the total number of detected photons does not depend on the time profile of the ALP injection, although the normalization factor $N_{o}$ does. Motivated by the studies of \cite{2007PhRvL..99w1102H} and \cite{2008PhLB..659..847D}, the values of $m_{a}$ and $g_{{\rm a}\gamma}$ are taken from $10^{-5}$ to $10^{-8}\,\rm{eV}$ and from $3.16\times10^{-13}$ to $10^{-10}\,\rm{GeV}^{-1}$, respectively. 

Then, we impose the following conditions on the number of observed photons:

\begin{enumerate}
    \item At least 0.5 photons have energies between $10 - 25\,\rm{TeV}$.
    \item Less than 0.5 photons have energies $> 25\,\rm{TeV}$.
\end{enumerate}

\begin{figure}[ht!]
    \centering
    \includegraphics[width=0.85\textwidth]{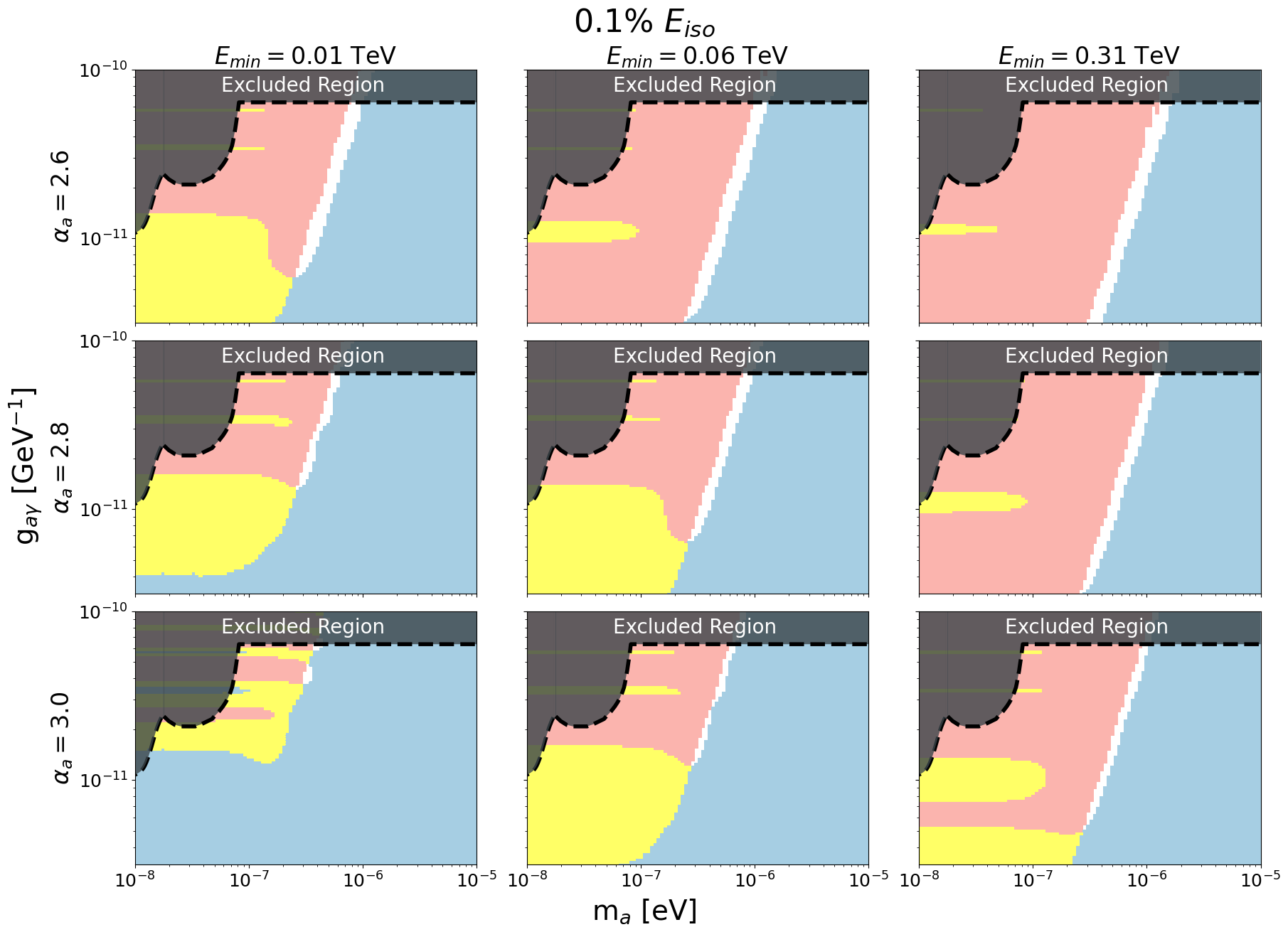}
    \caption{ALPs parameter space taking different values of $E_{\rm{min}}$ and $\alpha_{\rm a}$ for  $0.1\%$ of the $E_{\rm{iso}}$ taken by the ALPs burst. Pink: region with $>0.5$ photons in the energy range $10-25$ TeV. Blues: region with $<0.5$ photons for energies above $25~\text{TeV}$. Yellow: Allowed region obtained from the intersection of the pink and blue regions. The excluded region covers those reported by \cite{2016PhRvL.116p1101A}, \cite{2017NatPh..13..584A} and \cite{2019A&A...627A.159H}. As the minimum integration energy increases, the allowed region of candidates tends to favor softer ALP spectral indices.}
    \label{fig:0.01Eiso}
\end{figure}

\begin{figure}[ht!]
    \centering
    \includegraphics[width=0.7\textwidth]{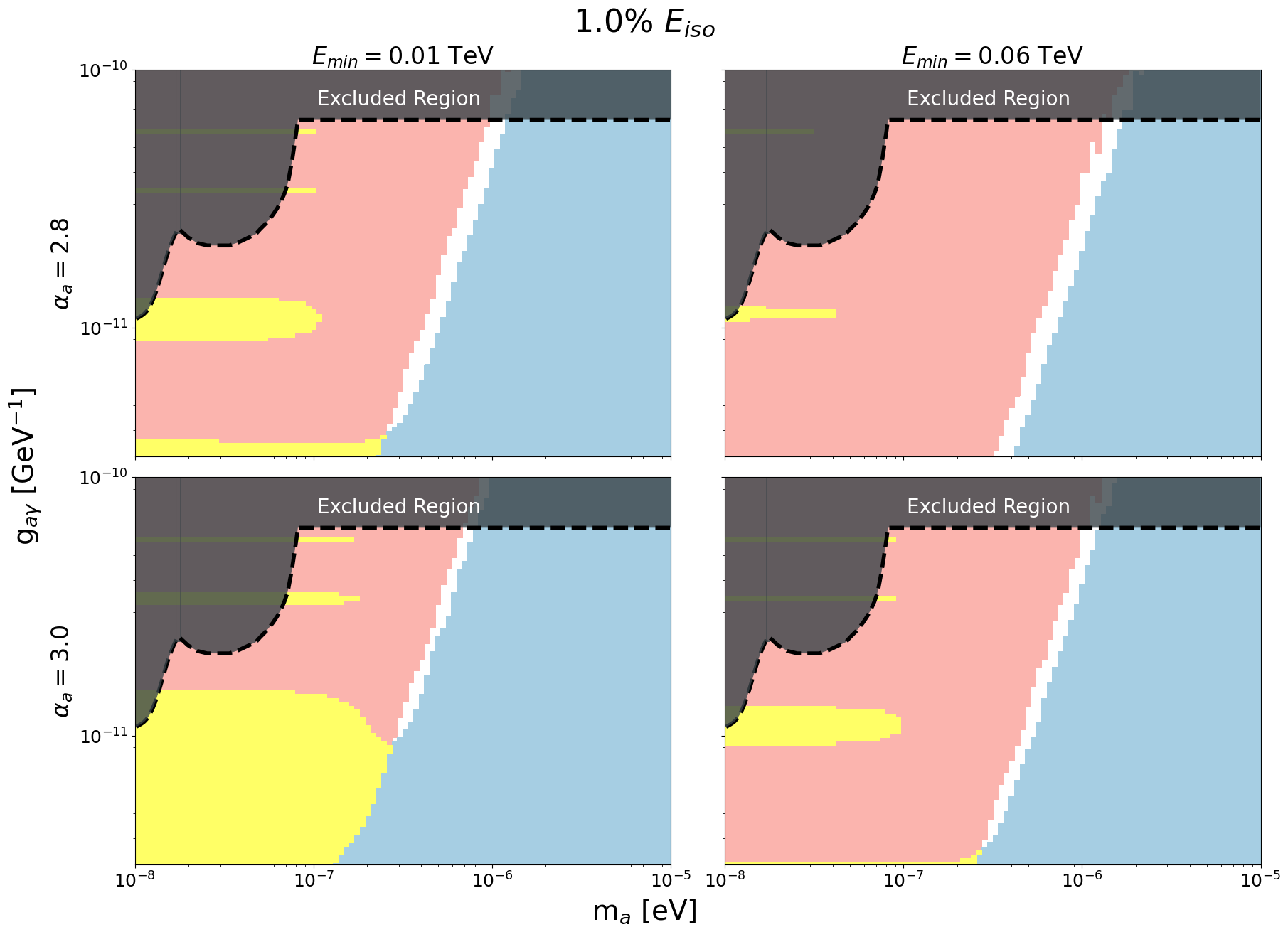}
    \caption{ALPs parameter space taking different values of $E_{\rm{min}}$ and $\alpha_{\rm a}$ for  $1\%$ of the $E_{\rm{iso}}$ taken by the ALPs burst. The color code is the same as in Figure \ref{fig:0.01Eiso}. Similarly to Figure \ref{fig:0.01Eiso}, as the minimum integration energy increases, the allowed region of candidates tends to favor softer ALP spectral indices. Additionally, the allowed region decreases as a larger amount of energy is taken by ALPs.}
    \label{fig:0.10Eiso}
\end{figure}

\begin{figure}[ht!]
    \centering
    \includegraphics[width=0.7\textwidth]{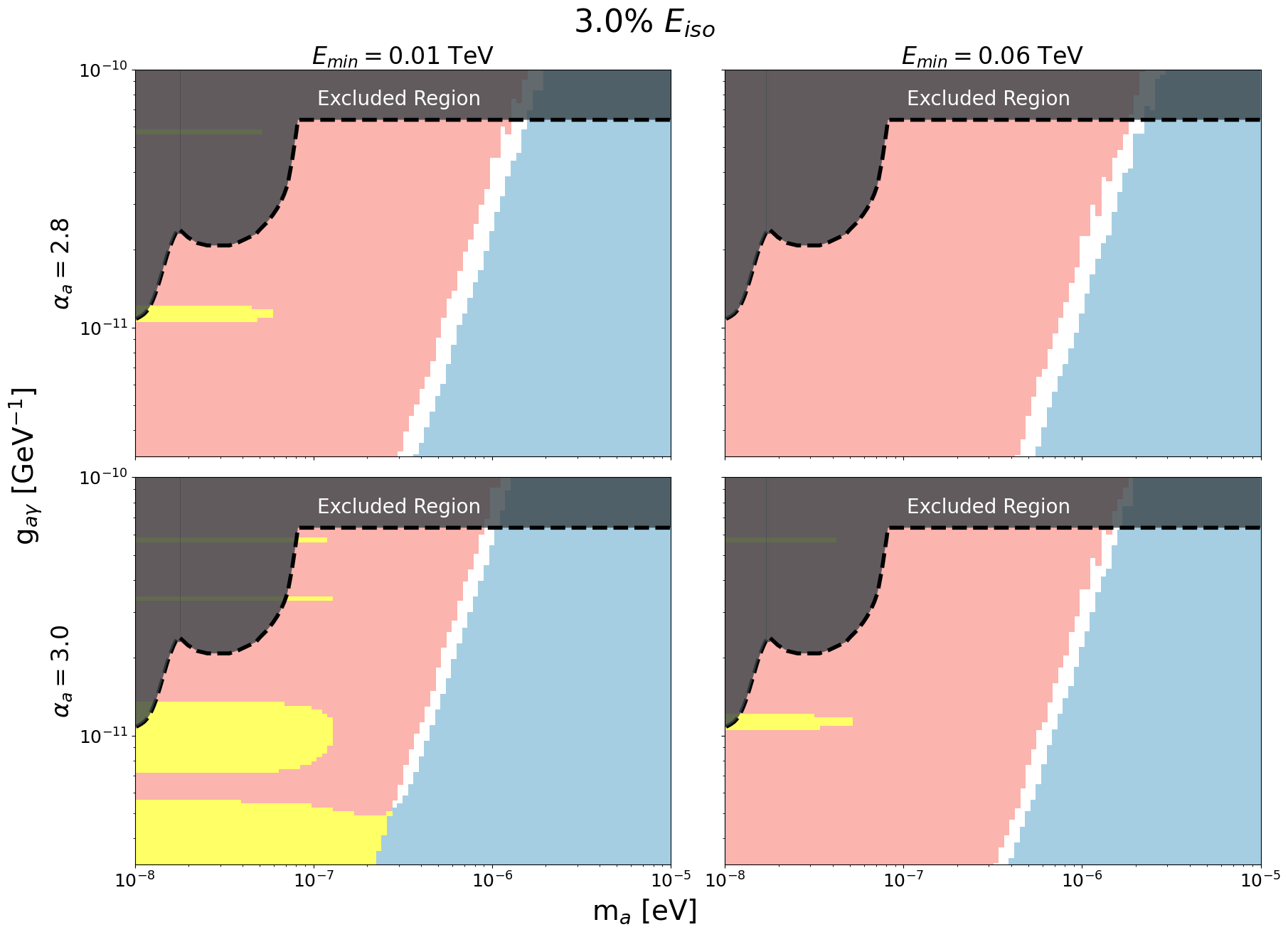}
    \caption{ALPs parameter space taking different values of $E_{\rm{min}}$ and $\alpha_{\rm a}$ for  $3\%$ of the $E_{\rm{iso}}$ taken by the ALPs burst. The color code is the same as in Figure \ref{fig:0.01Eiso}. The allowed region disappears for hard ALP spectra and high values of the minimum integration energy.}
    \label{fig:0.30Eiso}
\end{figure}

These conditions were set based on the fact that the highest-energy photon detected by LHAASO had an energy of $12.2^{+3.5}_{-2.4}\,\text{TeV}$, and no detection has been reported beyond the energy uncertainty band at this energy when a description of the spectra by a power law. We show the allowed parameter space for $\% E_{\rm{iso}}=0.1, 1 $ and $3$ in figure \ref{fig:0.01Eiso}, \ref{fig:0.10Eiso} and \ref{fig:0.30Eiso} assuming different values of $\alpha_{\rm a}$ and $E_{\rm min}$.  We observe that as the amount of energy carried by the ALPs increases, a softer spectrum and a lower $E_{\rm{min}}$ (tens of GeV or lower) are required to obtain a larger permitted region of candidates. Then, if an injected beam of ALPs by the GRB progenitor is responsible for the observed photons at 13 TeV, the total energy carried by ALPs must be a few percent of the isotropic energy of the GRB. Interestingly, it is still possible to find ALP candidates when the ALP spectra are harder than $2.6$ or when the minimum energy of the injected ALPs is above 1 TeV, they mainly lay within a region approximately from masses of $3\times10^{-7}$ and coupling factor of $2\times10^{-12}$ to masses of $1\times10^{-6}$ and coupling factors of $5\times10^{-11}$ as observed in figures \ref{fig:0.10Eiso} and \ref{fig:0.30Eiso}.

\begin{figure}[ht!]

    \includegraphics[width=0.49\textwidth]{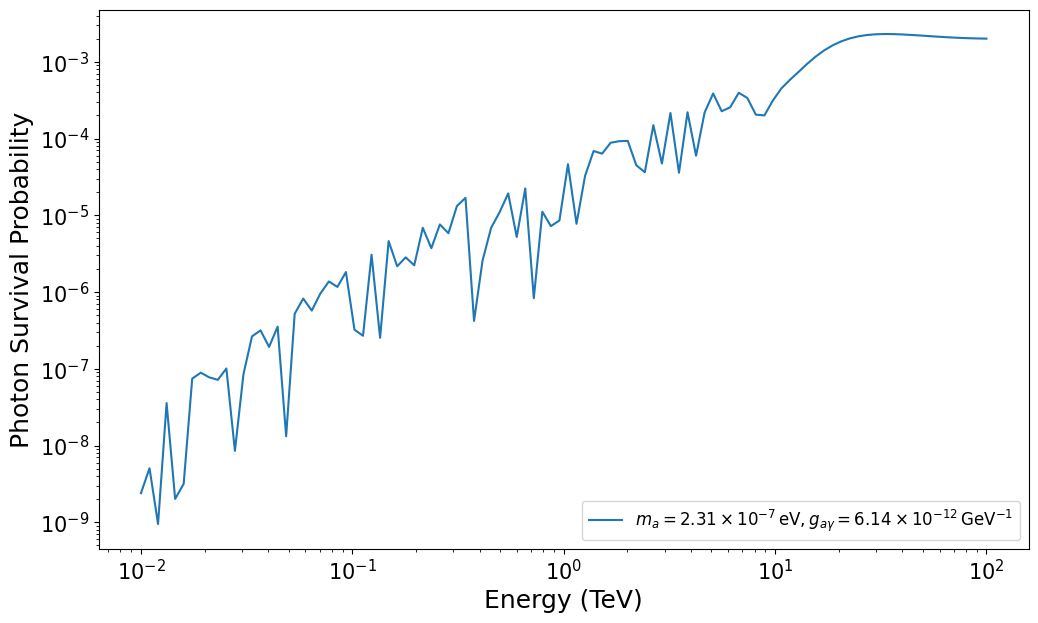}
    \includegraphics[width=0.49\textwidth]{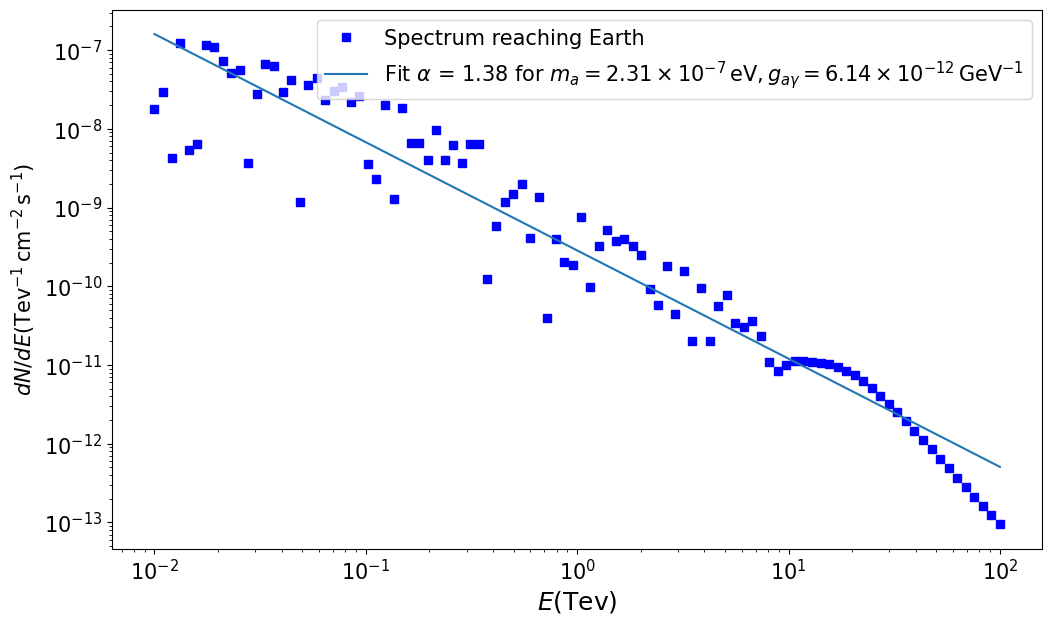}
    \caption{Left: Survival probability for a candidate to reach Earth as a photon. The value of the parameters for this ALP candidate are: $m_a=5.46\times10^{-7}~\text{eV}$ and $g_{a\gamma}=1.75\times10^{-11}~\text{GeV}^{-1}$. Right: Fitted photon spectrum for the candidate as in the left plot.}
    \label{fig:probs}
\end{figure}

\begin{figure}[ht!]
    \centering
    \includegraphics[width=0.85\textwidth]{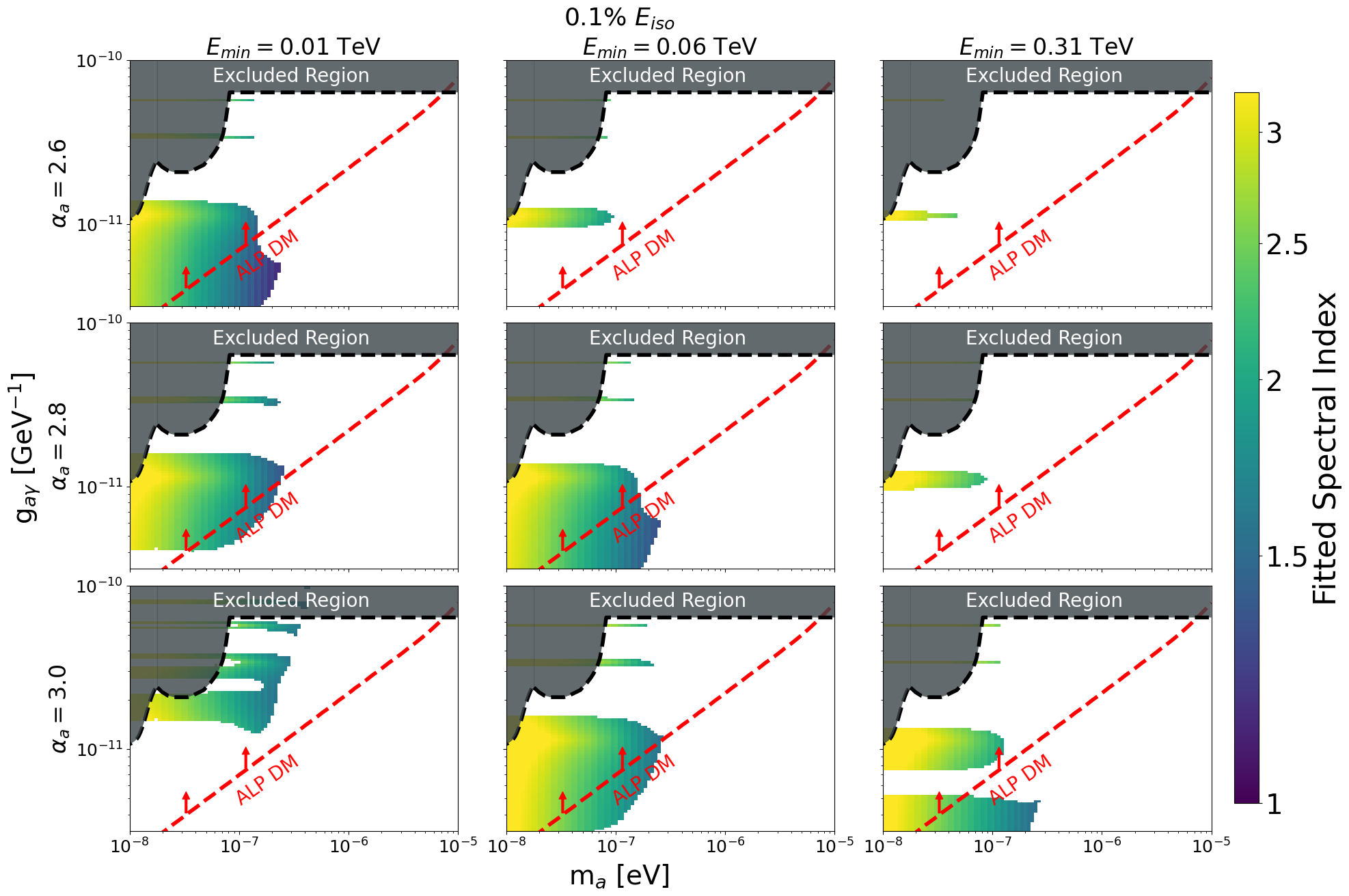}
    \caption{Fitted spectral indexes of the photon flux from candidates in the permitted region. Plots show results for $0.1\%E_{\rm{iso}}$. Candidates that reproduce hard photon spectra indices have masses between $0.2\times 10^{-7}$ and $3\times 10^{-7}$ regardless of the ALP spectral index or minimum integration energy.}
    \label{fig:indexmesh1}
\end{figure}

\begin{figure}[ht!]
    \centering
    \includegraphics[width=0.7\textwidth]{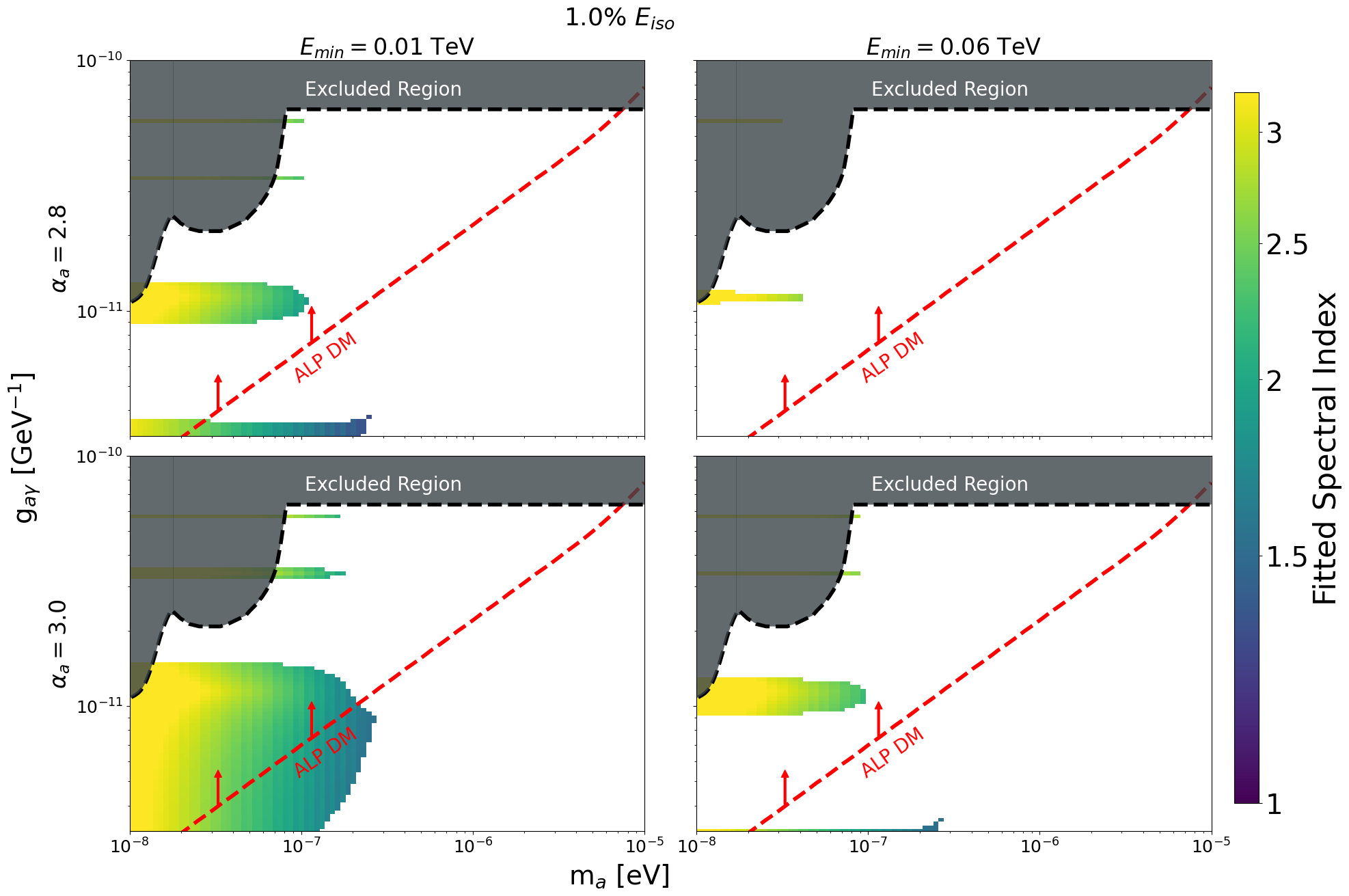}
    \caption{Fitted spectral indexes of the photon flux from candidates in the permitted region. Plots show results for $1\%E_{\rm{iso}}$. Candidates, if they exist, that reproduce hard photon spectra indices have masses between $0.2\times 10^{-7}$ and $3\times 10^{-7}$.}
    \label{fig:indexmesh2}
\end{figure}

\begin{figure}[ht!]
    \centering
    \includegraphics[width=0.7\textwidth]{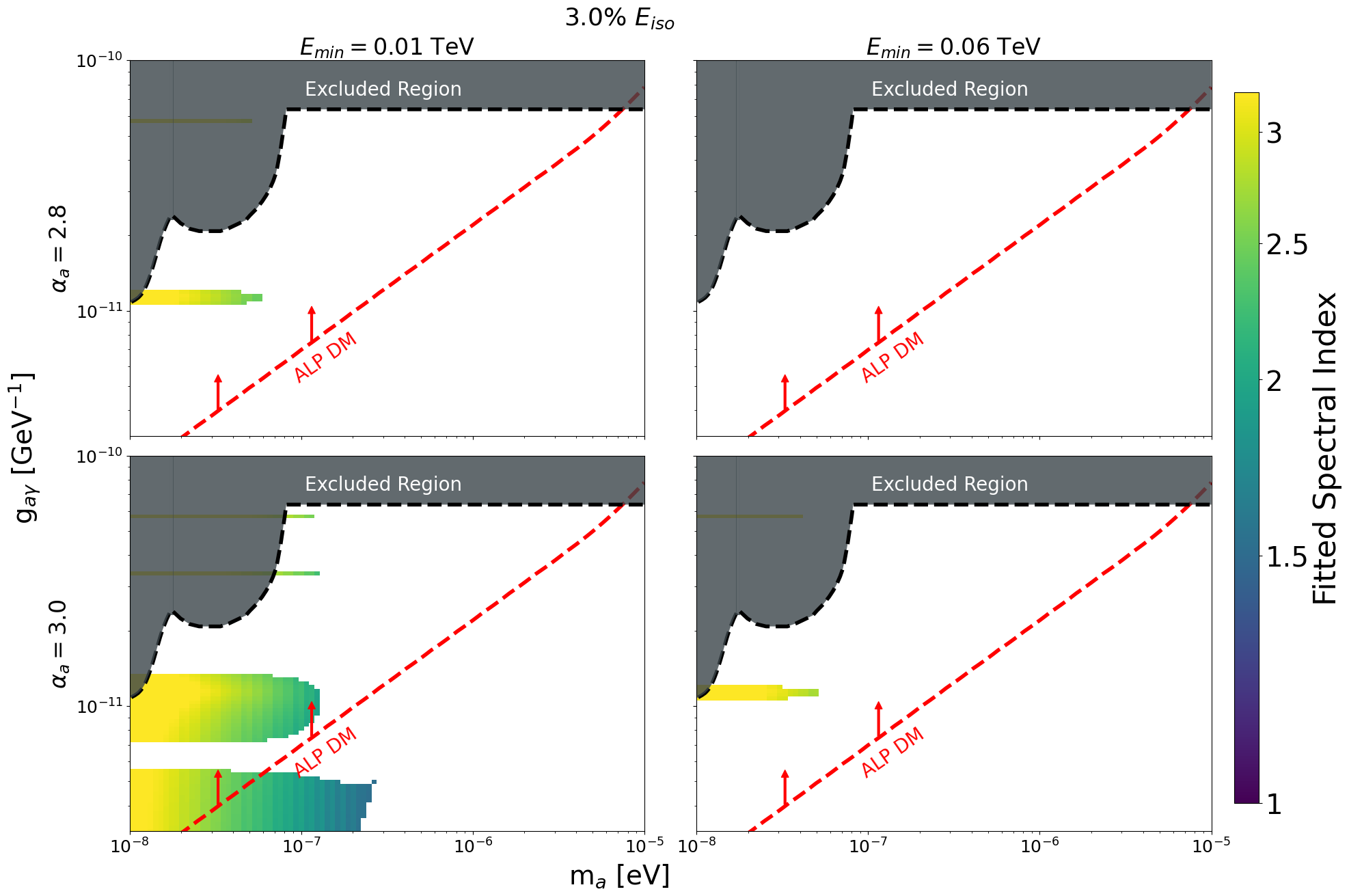}
    \caption{Fitted spectral indexes of the photon flux from candidates in the permitted region. Plots show results for $3\%E_{\rm{iso}}$. Candidates that reproduce hard photon spectral indices only exist with soft spectra and masses between $0.2\times 10^{-7}$ and $3\times 10^{-7}$.}
    \label{fig:indexmesh3}
\end{figure}

\begin{figure}[t!]
    \begin{minipage}{\textwidth}
        \centering
        \includegraphics[width=0.76\textwidth]{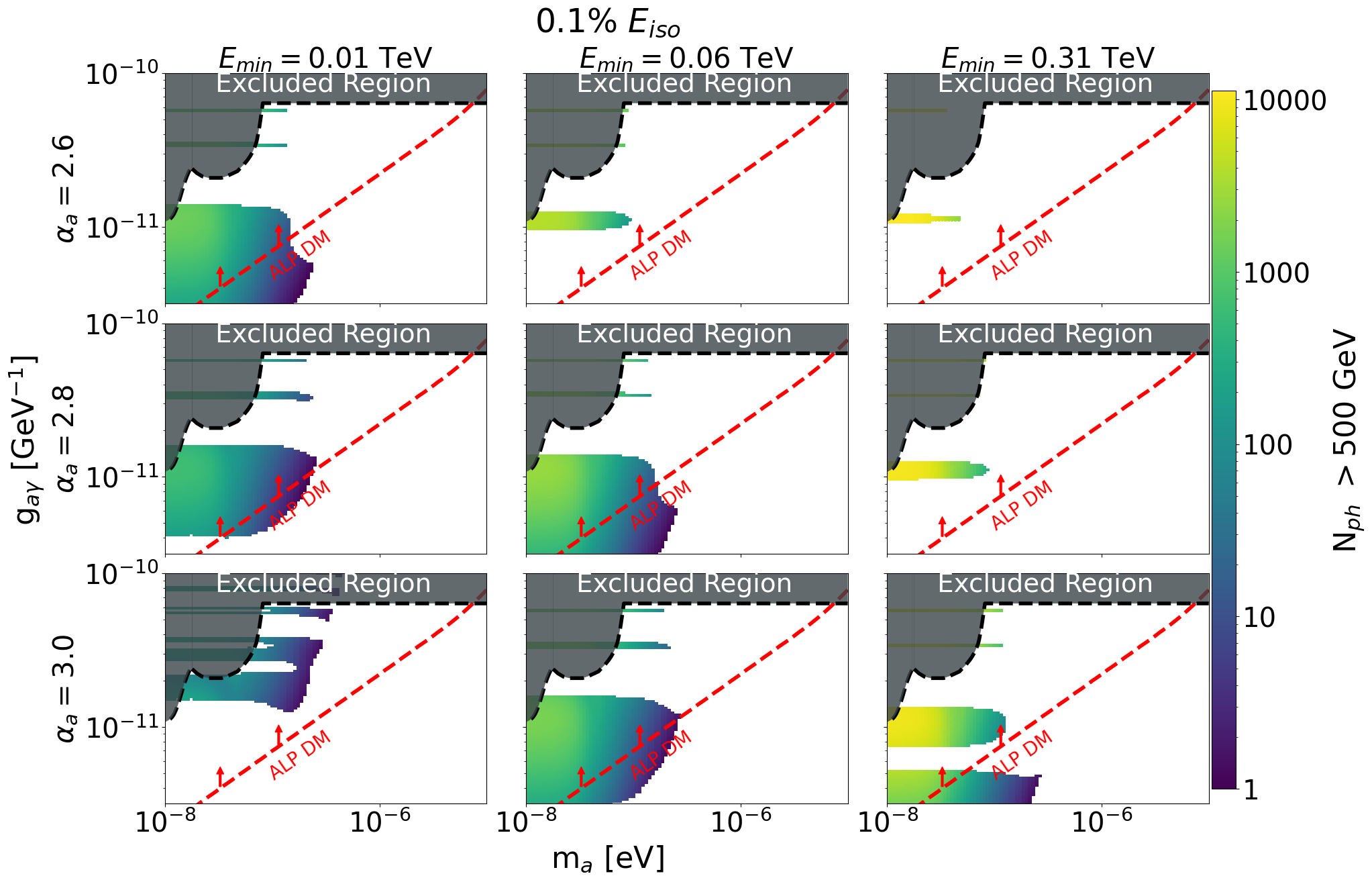}
    \end{minipage}\par
    \begin{minipage}{\textwidth}
        \centering
        \includegraphics[width=0.76\textwidth]{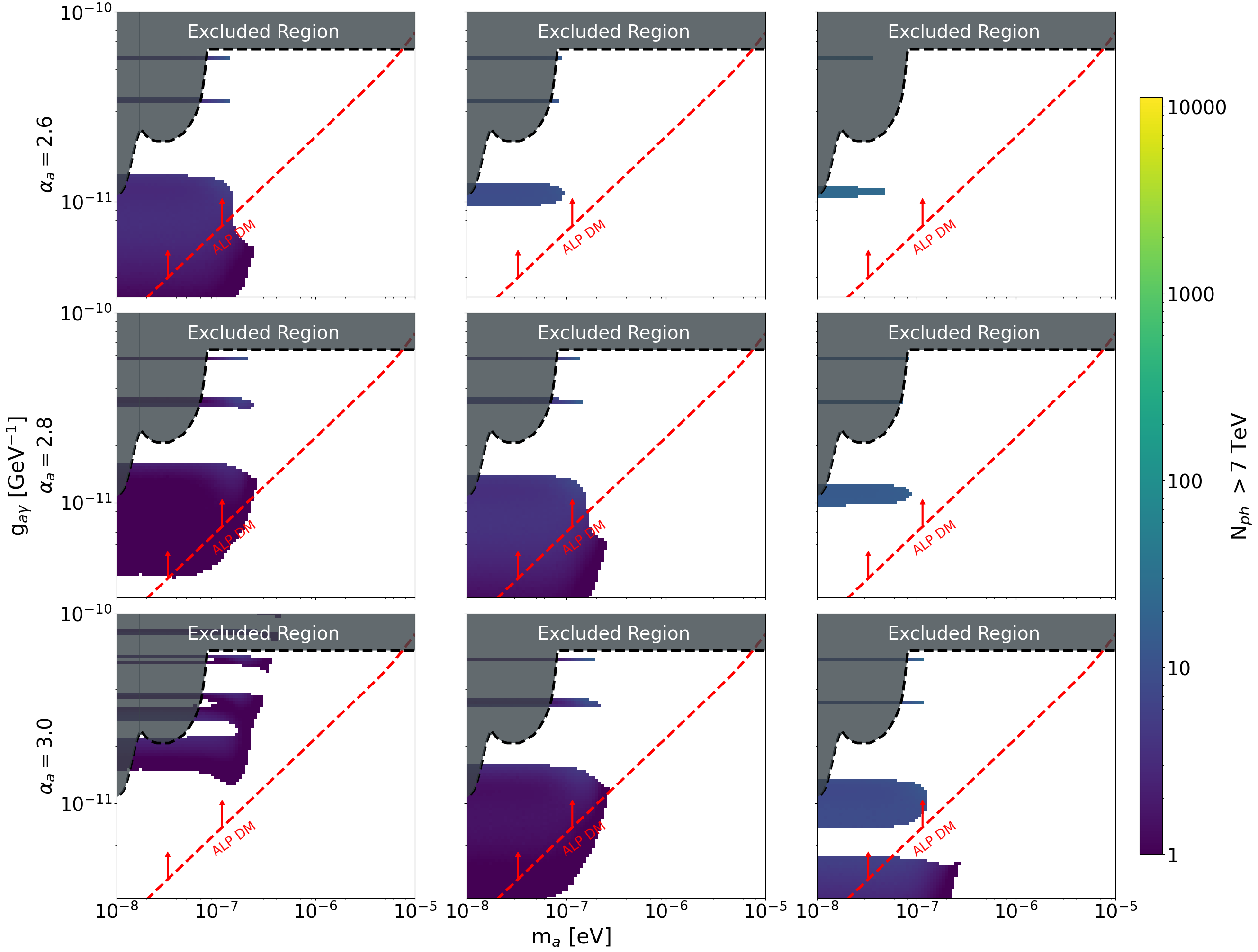}
    \end{minipage}\par
    \caption{Three upper rows: Number of photons above $500\,\rm{GeV}$ for ALPs candidates in the permitted region. Three lower rows: Number of photons above $7\,\rm{TeV}$ for ALPs candidates in the permitted region. Plots show results for $0.1\%E_{\rm{iso}}$ taken for the ALPs production, and different values of $\alpha_{\rm a}$ and $E_{\rm{min}}$. The figure demonstrates that there are multiple candidates capable of explaining the detection of a 13 TeV photon while contributing minimally to the number of photons detected between 500 GeV and 7 TeV.}
    \label{fig:8}
\end{figure}

To determine whether the allowed ALP candidates can reproduce the observed photon flux, we calculate the survival probability for each candidate and fit the resulting photon spectrum, $\alpha_{\gamma}$, that reaches Earth. Figure \ref{fig:probs} shows the survival probability as well as the fitted observed photon spectrum for the candidate defined by $g_{\rm a \gamma}=1.75 \times 10^{-11} {\rm GeV^{-1}}$, $m_a =5.46 \times 10^{-7} {\rm eV}$ with $\%E_{\rm{iso}} = 0.1$, $E_{\rm{min}}=31$ GeV, and $\alpha_{\rm a}=3.0$. For this candidate, the survival probability increases with energy, and a soft ALP spectrum is required to avoid observing photons beyond energies greater than 25 TeV. The resulting photon flux is characterized by a hard photon index of 2.15. We repeated this analysis for every permitted ALP candidate and show the resulting photon index, considering $\%E_{\rm{iso}}$ values of 0.1, 1, and 3 percent for different values of $E_{\rm{min}}$ and $\alpha_{\rm a}$ (figures \ref{fig:indexmesh1}, \ref{fig:indexmesh2} and \ref{fig:indexmesh3}). Then we calculated the number of photons with energies above $500\,\rm{GeV}$ that LHAASO would detect from these ALPs candidates, some of the results are shown in figure \ref{fig:8}. In general, we observed that there are several candidates capable of explaining the detection of a $13\,\rm{TeV}$ photon while making only a minimal contribution to the number of photons detected between 500 GeV and 7 TeV. This indicates that the ALPs scenario does not conflict at all with the reported results for this GRB at energies below 7 TeV by LHASSO and Fermi.

We also found that ALP candidates that reproduce hard photon spectra ($\alpha_{\gamma} \lesssim 2.26$) can only account for less than a dozen of photons with energies exceeding 500 GeV. Therefore, in order to explain a significant flux of photons with $E>500$ GeV, either the observed photon spectra must be extremely soft ($\gtrsim 3$), or the photons must originate from other process, such as the SSC mechanism, in line with LHAASO results up to $7\,\rm{TeV}$. On the other hand, if an injected beam of ALPs from the GRB progenitor is responsible for the observed 13 TeV photons, their spectra must be harder than the SSC contribution to explain the additional spectral component that LHAASO suggested. Therefore, candidates generating soft photon spectra (e.g., $\gtrsim 2.26 \pm 0.02$ with the Saldana-Lopez EBL model as reported by LHAASO) are not preferred. 

It is worth noting that in all studied cases, there exist a region characterized by hard spectra with spectral indices around 2. This holds true independently of the ALP spectrum and the percentage of energy carried by the ALP population. These candidates also contribute to the photon flux with one or two photons with energies above 7 TeV as shown in figure \ref{fig:8}. Finally, it is remarkable that the most likely ALP candidates for explaining an extra spectral component above 7 TeV are found below the dashed red line in Figures \ref{fig:indexmesh1}, \ref{fig:indexmesh2} and \ref{fig:indexmesh3}. Below this line, ALPs candidates could account for all the cold dark matter in the Universe \cite{2012JCAP...06..013A}.

\begin{figure}[ht!]
    \centering
    \includegraphics[width=0.7\textwidth]{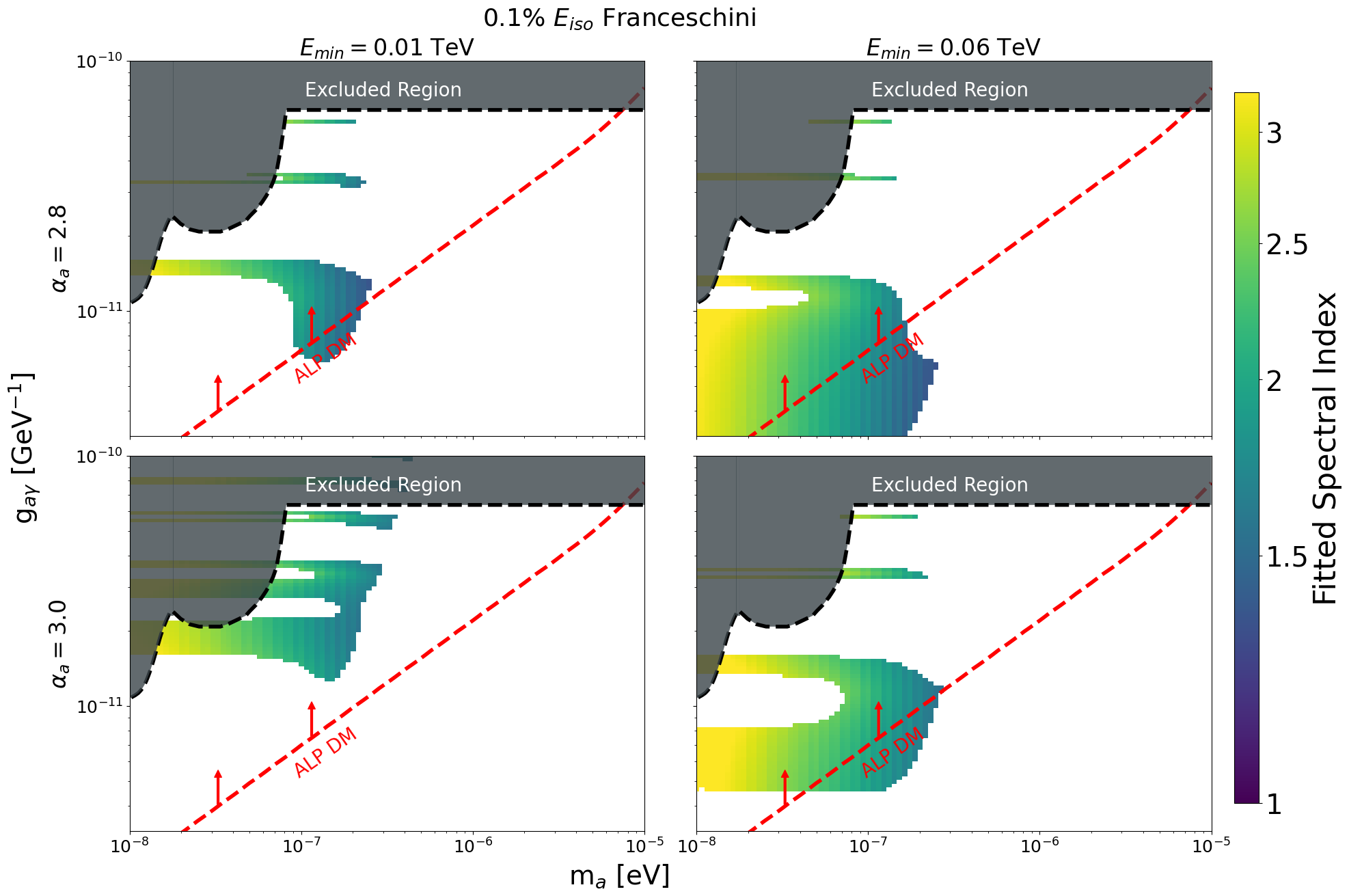}\\
    \includegraphics[width=0.7\textwidth]{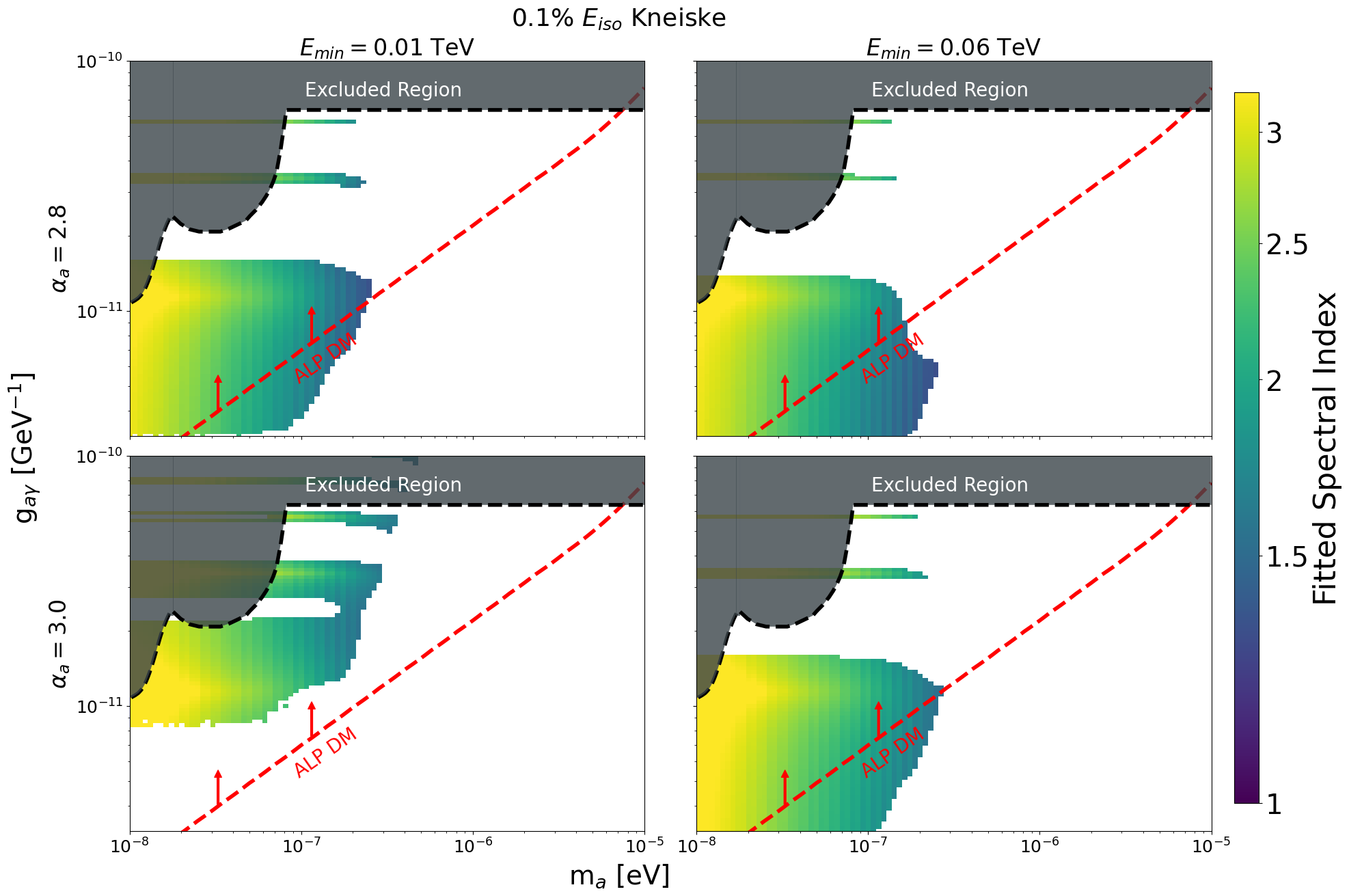}
    \caption{Fitted spectral indexes of the photon flux from candidates in the permitted region assiming the Francechini and Kneiske EBL models. Plots show results for $0.1\%E_{\rm{iso}}$. Candidates that reproduce hard photon spectral indices only exist with soft spectra and masses between $0.2\times 10^{-7}$ and $3\times 10^{-7}$ as obtained with Gilmore EBL model.}
    \label{fig:frakneindex}
\end{figure}

Finally, we perform the same analysis described above but using different EBL models to investigate the systematic uncertainties arising from the assumed EBL model. We focus on the most extreme EBL models: \cite{2010A&A...515A..19K} and \cite{2017A&A...603A..34F}. The attenuation of photons due to the EBL is the largest for the Francheschini model, moderate for the Gilmore model, and the smallest among of all models for the Kneiske model. We compute the allowed regions and compare them with those obtained using the Gilmore EBL model. The region of allowed candidates according to the Kneiske EBL model encompasses those allowed by both the Gilmore and Francheschini models. Similarly, the region allowed by Gilmore model includes those allowed by Franceschini model. We find that the systematic uncertainties resulting from assuming different EBL models decrease as the percentage of isotropic energy of the GRB taken by the ALPs burst increases. Specifically, the differences reach up to 68\% and 52\% for two specific combinations of the ALPs spectral index and the minimum integration energy. For all other cases, the difference are below 15\%, and they are minimal (less than 1\%) when 3\% of the isotropic energy is taken by the ALPs population. For those cases where the systematic uncertainty is maximal, we calculate the fitted spectral indices of the photon flux from those candidates and present the results in \ref{fig:frakneindex}. As observed, regardless of the EBL model, there exist potential ALP candidates capable of explaining the TeV emission observed by LHAASO from GRB 221009A. The allowed candidates are generally cluster around masses of $10^{-7}$ eV. However, with greater attenuation, the lower coupling factors are discarded, in other words rejecting the region where ALPs could account for all the cold matter in the Universe. The most significant result is the existence of potential candidates capable of reproducing a hard photon spectrum as the observed by LHAASO.

\section{Final remarks}\label{sec:res}

We have analyzed the parameter space for an ALPs population released at the GRB that could potentially explain the detection of the $13\ \rm{TeV}$ photon by LHAASO. To narrow down the candidates, we set the condition that the number of photons between $10-25\ \rm{TeV}$ must be at least 0.5 photons, while the number of photons above $25\ \rm{TeV}$ must be less than 0.5 photons. Based on this and assuming the Gilmore EBL model, we calculated the resulting gamma-ray spectrum of the candidates in the permitted regions, where the fitted spectral indexes ranged between $1.8-3.0$. Then, we calculated the number of photons above $500\ \rm{GeV}$ and found that, for those candidates capable of explaining LHAASO's observation, the contribution does not exceed hundred of photons. The contribution to the photon flux from ALPs up to $7\ \rm{TeV}$ is less than $10\%$ of the photons reported by LHAASO, which favors the Synchrotron self-Compton scenario below this energy. The most favorable candidates to explain the extra component above $7\ \rm{TeV}$ are those that show a harder photon spectral index. We have considered several EBL models. A EBL model predicting larger attenuation tend to reject ALP candidates with the lowest coupling factor. However,  for some hypothesis of EBL model, these candidates are found below a region of the parameter space in which, if detected, ALPs could account for all of the cold dark matter in the Universe. This region will be accessible to future experiments such as IAXO \citep{2019JCAP...06..047A} and DMRadio-m3 \citep{2022PhRvD.106j3008B}, which their future results will help to unravel this paradigm.

The ALPs scenario explored in this work considers that the energy carry by the ALPs is a small fraction of the GRB energy, which is  $\leq3\%$. Moreover, the energy is carried by ALPs with energy as low as tens of GeV through a spectrum extending at least up to tens of TeV. This is consistent with previous studies \cite{2018IJMPD..2742003N} of GRB emission at hundred of GeV and up to 4 TeV where the Synchrotron self-Compton mechanisms describes the observations reasonably well. While the mechanisms of production and acceleration of light DM particles such as ALPs are still under investigation, it is plausible that a population of ALPs would be able to explain the 13-TeV photon and those candidates that would account for all DM of the Universe result in an extra spectral component harder than the SSC component for the gamma-ray flux.

This work was supported by UNAM-PAPIIT project number IG101323.

\bibliography{sample631}{}
\bibliographystyle{aasjournal}

\end{document}